\documentclass[conference]{IEEEtran}

\usepackage{amsmath,amssymb,amsfonts}
\usepackage{algorithmic}
\usepackage{cite}
\usepackage{csquotes}
\usepackage{enumerate}
\usepackage{graphicx}
\usepackage{subcaption}
\usepackage{textcomp}
\renewcommand{\theparagraph}{%
  \ifnum\value{subsubsection}=0%
    \thesubsection\alph{paragraph}%
  \else%
    \thesubsubsection\alph{paragraph}%
  \fi%
}

\usepackage[pdf]{dot2tex} %

\usepackage{pgfplots}
\usepgfplotslibrary{colorbrewer}
\pgfplotsset{
  cycle list/.define={my marks}{
    every mark/.append style={solid,fill=\pgfkeysvalueof{/pgfplots/mark list fill}},mark=*\\
    every mark/.append style={solid,fill=\pgfkeysvalueof{/pgfplots/mark list fill}},mark=square*\\
    every mark/.append style={solid,fill=\pgfkeysvalueof{/pgfplots/mark list fill}},mark=triangle*\\
    every mark/.append style={solid,fill=\pgfkeysvalueof{/pgfplots/mark list fill}},mark=diamond*\\
  },
}
\newcommand{\resultaxis}[1]{%
  \begin{tikzpicture}
    \footnotesize
    \begin{axis}[
      scale only axis,
      height=.80\linewidth,
      width=.85\linewidth,
      xtick={0.25, 0.3, 0.35, 0.4, 0.45},
      legend pos=north west,
      legend style={fill=none, draw=none, font=\strut},
      legend cell align={left},
      ymin=0.15,
      ymax=0.85,
      xmin=0.25,
      xmax=0.45,
      enlarge x limits=0.05,
      enlarge y limits=0.05,
      xlabel={$\alpha$},
      cycle list/Set1,
      mark list fill={.!75!white},
      cycle multiindex* list={
        Set1
        \nextlist
        my marks
        \nextlist
        linestyles
        \nextlist
        very thick
        \nextlist
      },
      ]
      #1
      \addplot[ultra thin, black, dashed, no markers, domain=0.25:0.45, forget plot] {x};
      \addplot[black, dotted, no markers, domain=0.25:0.45, forget plot] {x / (1-x)};
    \end{axis}
  \end{tikzpicture}
}

\usepackage[hidelinks]{hyperref}

\begin{document}

\title{Parallel Proof-of-Work with DAG-Style Voting\\and Targeted Reward Discounting} %

\author{Patrik Keller}

\maketitle

\begin{abstract}

  We present parallel proof-of-work with DAG-style voting, a novel proof-of-work cryptocurrency protocol that, compared to Bitcoin, provides better consistency guarantees, higher transaction throughput, lower transaction confirmation latency, and higher resilience against incentive attacks.
  The superior consistency guarantees follow from implementing parallel proof-of-work, a recent consensus scheme that enforces a configurable number of proof-of-work votes per block.
  Our work is inspired by another recent protocol, Tailstorm, which structures the individual votes as tree and mitigates incentive attacks by discounting the mining rewards proportionally to the depth of the tree.
  We propose to structure the votes as a directed acyclic graph (DAG) instead of a tree.
  This allows for a more targeted punishment of offending miners and, as we show through a reinforcement learning based attack search, makes the protocol even more resilient to incentive attacks.
  An interesting by-product of our analysis is that parallel proof-of-work without reward discounting is less resilient to incentive attacks than Bitcoin in some realistic network scenarios.

\end{abstract}


\section{Introduction}\label{sec:intro}

Proof-of-work cryptocurrenies expose a circular dependency among the consensus algorithm and the incentive scheme:
miners invest computational resources to establish consensus about a growing list of cryptocurrency transactions
and the incentive scheme awards miners with cryptocurrency-denominated rewards to motivate their participation.
Analysing the consistency guarantees of cryptocurrency protocols is vital as inconsistencies enable double spending attacks.
But the incentives matter just as much:
guaranteeing consistency is only possible if a majority of miners follows the protocol;
misaligned incentives motivate dishonest behaviour, turn rational miners malicious, and ultimately threaten consistency.

To date, we are well informed about the security properties of Bitcoin.
The cryptocurrency guarantees eventual consistency~\cite{GarayKL15, PassSS17sec, KifferRS18, DemboKTTVWZ20, GaziKR20}, and we also know the residual failure probability after waiting for a given time or number of confirmations~\cite{LiG021, Gazi0R22, Guo022}.
On the incentive side, we know optimal attack strategies~\cite{EyalS14, SapirshteinSZ16, GervaisKWGRC16}.
These strategies demonstrate that dishonest miners with more than a third of the hashrate can steal rewards from honest participants; attackers who are able to reorder messages can benefit independent of their hashrate.

Throughout the years, we have seen many protocol proposals claiming to improve on Bitcoin with respect to transaction confirmation latency and throughput.
Many proposals seem promising, but unfortunately, their analysis is not as thorough as Bitcoin's.
Usually, the proposers focus only on eventual consistency or incentives, rarely both.
Even more scarce are protocols that provide upper bounds for the failure probability after waiting for a given time or number of confirmations.
This is not surprising as is took about 14 years to complete these analyses for the relatively simple Bitcoin protocol.

We are aware of one recent exception, Tailstorm~\cite{KellerGBG23}, which improves on Bitcoin with respect to transaction confirmation latency, transaction throughput, and also resilience against incentive attacks.
Tailstorm implements parallel proof-of-work~\cite{KellerB22}, a recent consensus protocol with superior consistency guarantees:
after a 10-minute confirmation time, the probability of a successful double spend is approximately 50 times lower than in Bitcoin, given realistic network assumptions.
The gist of parallel proof-of-work is that it requires a configurable number of proof-of-work votes, before appending the next block becomes possible.
To increase transaction throughput, Tailstorm records transactions in each vote.
To reduce confirmation latency, Tailstorm structures the votes as tree: votes mined later can confirm votes mined earlier.
The authors observe that deep trees imply high linearity and more transactions ordered.
Wide trees, on the other hand, can be caused by network delays or dishonest behavior.
Based on this observation, Tailstorm discounts rewards proportional to the depth of the tree.
As it turns out, this reward discounting makes the protocol more resilient to incentive attacks.

We observe two shortcomings in Tailstorm.
First, as the votes are structured in a tree, miners have to decide which branch they extend.
This implies that some votes remain unconfirmed until the next block.
Hence, not all transactions benefit from the lower confirmation latency.
Second, as the reward discounting applies uniformly to all votes of each tree, some miners will get punished even if they did not cause any non-linearity themselves.
Based on these observations, we make the following contributions:

\begin{enumerate}
  \item We guide the reader through the incremental development of parallel proof-of-work, tree-style voting, and depth-based reward discounting. On the way, we simplify the original protocols considerably.
  \item We propose a new proof-of-work consensus protocol that structures the votes as directed acyclic graph (DAG) to enable lower transaction confirmation latency in the pessimistic case, while preserving the superior consistency guarantees of parallel proof-of-work.
  \item We propose a targeted reward discounting scheme that punishes only those miners who contribute to the non-linear parts of the DAG.
  \item We employ reinforcement learning to search for effective incentive attacks against Bitcoin, parallel proof-of-work, tree-style voting with depth-based discounting and our proposed protocol.
    Our results support the following conclusions.
    \begin{enumerate}
      \item
        Parallel proof-of-work without discounting is less resilient to incentive attacks than Bitcoin.
        This contradicts earlier claims~\cite{KellerGBG23}, which attributed the problem to the leadership election employed in the original parallel proof-of-work protocol~\cite{KellerB22}, but not here.
      \item
        Tree-style voting with depth-based discounting is more resilient to incentive attacks than Bitcoin.
      \item
        The proposed DAG-style voting with targeted reward discounting is even more resilient to incentive attacks.
    \end{enumerate}
\end{enumerate}

We structure the paper along these contributions.
We start by describing our system model in Section~\ref{sec:model}.
We then describe the four mentioned protocols:
Bitcoin in Section~\ref{sec:seq},
parallel proof-of-work in Section~\ref{sec:par},
tree-style voting in Section~\ref{sec:tree}, and
DAG-style voting in Section~\ref{sec:dag}.
Section~\ref{sec:eval} presents our evaluation methodology and results.
We quickly describe how to configure our proposed protocol for deployment in Section~\ref{sec:config}.
We discuss our results in Section~\ref{sec:discussion} and conclude in Section~\ref{sec:conclusion}.
\section{System Model}\label{sec:model}

\paragraph{Participants}
We see proof-of-work blockchains as distributed systems, where the participating \emph{nodes} manage their own local state and interact by exchanging \emph{messages} across a pre-established peer-to-peer network.
In the context of blockchain systems, it is useful to think of each node as one physical machine connected to the internet.
The \emph{protocol} specifies the nodes' behavior.
More specifically, the protocol prescribes how nodes update their state and what messages they send in reaction to the messages received.

Each node has an \emph{operator}.
We distinguish two types of operators, \emph{attackers} and \emph{defenders}.
Defenders instruct their nodes to follow the protocol.
We call these nodes \emph{honest} or \emph{benign}.
Attackers \emph{may} instruct their nodes to disobey the protocol.
We call such nodes \emph{malicious} or \emph{Byzantine}.

\paragraph{Blockchain}
In practice, nodes manipulate, send, and receive binary data.
Through this binary lens, blockchains are hash-linked blobs of data.
Blobs can refer to other blobs by including a hash of the referenced blob.
The hash functions in use are practically collision-free, which makes the data structure write-only.
It is always possible to append a new blob.
Any change to a blob, however, also changes its hash:
any blob referring to the old version still includes the old hash and thus refers to the unmodified blob.

In theory, we do not want to deal with binary data and thus introduce an abstraction for blobs and hash-linking.
We reuse the prevalent terminology.
Blobs are now \emph{blocks}.
Each block has an arbitrary number of \emph{parent} blocks, including none. %
If block~$A$ is a parent of block~$B$, then block~$B$ is a \emph{child} of block~$A$.
Blocks that can be reached with the parent relationship from block~$A$ are called \emph{ancestors} of~$A$, and these that can be reached with the child relationship are called \emph{descendants} of~$A$.
Blocks may store arbitrary data in their \emph{body}.
Blocks, their parents, and their body are persistent.

The parent relationship spans a \emph{directed acyclic graph} (DAG).
Within the DAG, blocks without parents are called \emph{roots}.
Blocks without children are called \emph{leaves}.
We regularly use the \emph{topological ordering} of blocks within the DAG, where parents come before their children, but children of the same parents, \emph{siblings}, have the same rank.

Each block also defines a \emph{blockchain}, namely the block itself and all its ancestors.
Blockchains have exactly one leave, which we sometimes refer to as tip of the chain.

\paragraph{Proof-of-Work}
All protocols in this paper have in common, that appending any new block requires to solve a moderately hard puzzle.
In practice, the protocol would set a \emph{difficulty} threshold on the hash references.
Searching for valid blocks is called \emph{mining} and amounts to iteratively tweaking the blob and evaluating the hash-function until the threshold is met.
The node who solves the puzzle for a block is the \emph{miner} of this block.
The limiting factor is the \emph{hashrate} at which the nodes can evaluate the hash function.
This rate may change over time.
To maintain a somewhat regular block rate, the protocols employ a \emph{difficulty adjustment algorithm} (DAA).

In our model, we abstract from the underlying puzzle implementation.
We assume that the random time between any two consecutive blocks follows an exponential distribution with \emph{mining rate}~$\lambda$.
The \emph{expected block interval} is~$\lambda^{-1}$.

\paragraph{Communication}
Nodes send and receive blocks using a peer-to-peer broadcast network.
The communication is subject to network delays.
Initially, freshly mined blocks are only known to their miner.
The miner may decide to \emph{withhold} the block before \emph{releasing} it to the other nodes.
Honest miners typically share their blocks immediately.

In practice, invalid blocks are ignored on the broadcast layer.
Verifying a block, however, requires full knowledge of all ancestors.
This is because it is not possible to distinguish a link that was intentionally broken, e.\,g., by replacing the hash with a random number, from a link to an existing but not yet received block.
Throughout this paper we assume that nodes receive blocks in topological order, which drastically simplifies our protocol descriptions.

\section{Protocols}\label{sec:proto}

We proceed with describing four proof-work consensus protocols with their respective reward schemes.
We start with sequential proof-of-work as it is deployed in Bitcoin.
We then guide the reader step-by-step through the development of parallel proof-of-work with DAG-style voting.
This section focuses on intuitive justifications and existing results. Our own evidence follows in Section~\ref{sec:eval}.
We provide the full protocol specifications online~\cite{trainingrepo}.

\subsection{Sequential Proof-of-Work}\label{sec:seq}

\begin{figure}
  \centering
  \input{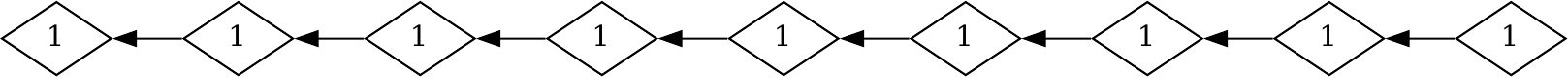}
  \caption{
    Example chain for sequential proof-of-work.
    Each diamond represents one block.
    Arrows indicate hash-references.
    The labels show the constant incentive scheme: each proof-of-work gets the same reward.
  }\label{fig:seq}
\end{figure}

Bitcoin miners build a linear chain of blocks.
Each block has exactly one parent and the miners always extend the longest chain they know.
The difficulty adjustment algorithm targets a chain growth rate of 6 blocks per hour.
Bitcoin's reward scheme assigns one unit of reward to (the miner of) each block.
Figure~\ref{fig:seq} shows an example chain, where each diamond represents one block and the arrows indicate the parent relationship.
The labels show the rewards.

The strictly linear chain structure implies that all blocks in the chain have been mined sequentially.
We thus refer to this protocol as sequential proof-of-work.
We want to highlight, that of any two blocks with the same parents, only one will be retained, the others are \emph{orphaned}.
This is particularly problematic when an attacker engages in a race for the longest chain against the defenders.
In worst-case scenarios, we have to assume that the attacker can induce network delays and thereby orphan some of the defenders' blocks.
This effectively slows down the defenders' mining and ultimately has the same effect as assigning more hashrate to the attacker.
The problem can be mitigated, however, by deviating from the linear chain structure.

\subsection{Parallel Proof-of-Work}\label{sec:par}

Parallel proof-of-work distinguishes between blocks and votes.
Appending a new block requires $k - 1$ votes confirming the previous block.
Together with the proof-of-work required for the block itself, this makes $k$~proofs-of-work per block.
The \emph{votes} do not depend on each other and hence can be mined in \emph{parallel}.
The \emph{blocks}, however, still form a \emph{linear} chain, similar to sequential proof-of-work.
Temporary ambiguities arise when there are two blocks confirming the same parent block.
Such conflicts are resolved quickly, by instructing the miners to vote for the block that has the most votes already.

\begin{figure}
  \centering
  \input{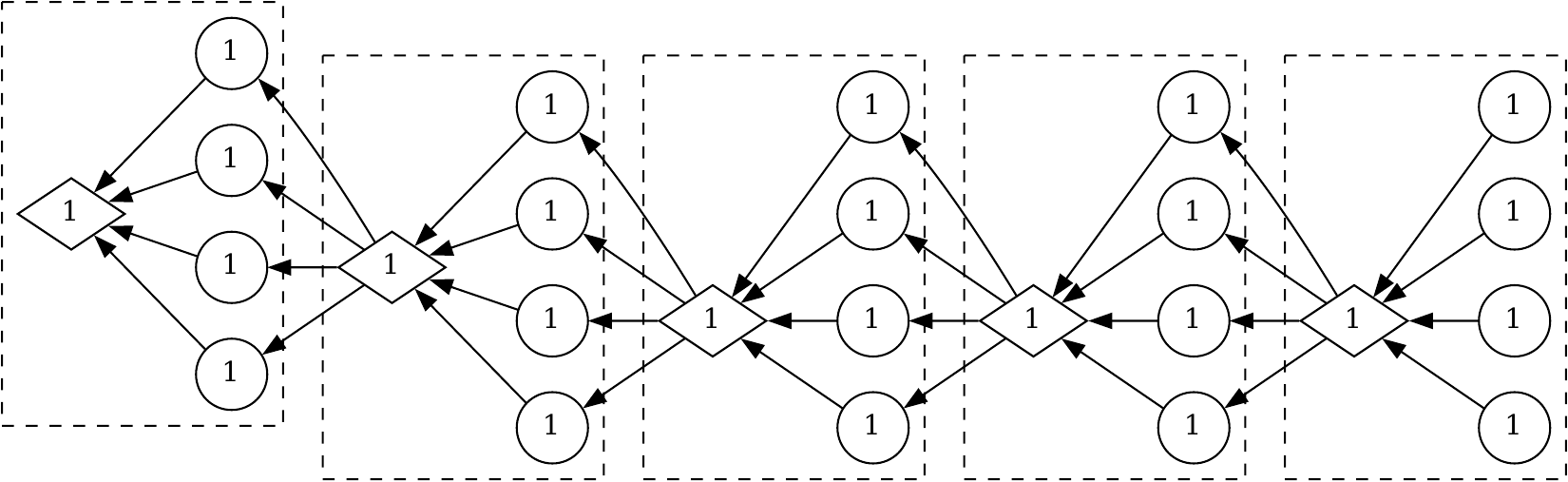}
  \caption{
    Example chain for parallel proof-of-work with $k = 5$ proofs-of-work per block.
    Diamonds represent blocks and circles represent votes.
    Arrows indicate hash-references.
    Dashed boxes mark epochs.
    The labels show the constant incentive scheme where each proof-of-work gets the same reward.
  }\label{fig:par}
\end{figure}

Parallel proof-of-work's reward scheme is constant; it assigns one unit of reward per proof-of-work, just like sequential proof-of-work does.
Figure~\ref{fig:par} shows an example chain for parallel proof-of-work with $k = 5$.
As before, blocks are represented by diamonds, arrows indicate hash references, and the reward is on the labels.
Votes are drawn as circles.
We also add dashed rectangles to mark the individual periods of parallelism, which we refer to as \emph{epochs}.
Observe that each block starts a new epoch, and during each epoch, the blockchain grows by $k$~proofs-of-work.

The main advantage of parallel proof-of-work is that votes confirming the same parent block are eligible for the next block, even when their miners suffer from communication delays.
This reduces the number of orphans and avoids slowing down the group of defenders in a block race.
As Keller and Böhme~\cite{KellerB22} argue, parallel proof-of-work implies lower residual consistency failure probabilities within the same time.

We want to highlight, though, that our version of parallel proof-of-work slightly differs from the original version presented at AFT\,'22~\cite{KellerB22}.
In the original version, blocks do not require a proof-of-work; only the votes do.
A priori, all nodes can propose their own block as soon as there are $k$ votes.
To avoid network congestion, the AFT\,'22 version employs a random leader election mechanism: only the miner of the vote with the smallest hash value is allowed to propose the next block.
According to Keller et al.~\cite{KellerGBG23}, this leader election mechanism makes the protocol more prone to incentive attacks than Bitcoin, particularly in optimistic network conditions.
Keller et  al.~propose to eliminate the leader election process and delaying the transmission of the block until the first confirming vote was mined~\cite{KellerGBG23}.
This is the approach we take here.
Additionally, we simplify the protocol by merging the block with the first confirming vote.

Structuring the $k$ proofs-of-work in parallel is good for consistency, but it can be seen as a shortcoming for the cryptocurrency, where the main goal is to order transactions.
Parallel proof-of-work writes the transactions into the blocks, not the votes, as only the blocks are structured linearly.
Writing the transactions into the votes might boost throughput, but it will not contribute to their ordering.
Tailstorm~\cite{KellerGBG23} leverages this observations and introduces a hybrid approach where the votes are structured linearly when possible, and fork out into a tree when necessary.

\subsection{Parallel Proof-of-Work with Tree-Style Voting}\label{sec:tree}

We present tree-style voting as an add-on to our simplified version of parallel proof-of-work described above.
There are still $k$ proofs-of-work per epoch.
Assembling the next block requires $k-1$ votes confirming the last.
All votes confirming the same parent block are eligible for the next block, enabling parallel mining of votes.
Miners resolve block conflicts by confirming the block most voted for.

The new aspect is that votes can now refer to another vote of the same epoch.
This results in a tree structure, where the last block is the root.
Within an epoch, miners append their vote to the longest branch they know.
This makes tree-style voting a hybrid between parallel and sequential proof-of-work.
In the optimistic case, if all votes are mined in sequence, the tree has a single branch, and the chain is linear.
In the event of network delays, parallel votes induce forks in the tree.
However, since all branches can be merged into the next block,
orphans are avoided in any case, and the superior consistency guarantees of parallel proof-of-work translate to tree-style voting.

One benefit of tree-style voting arises when storing cryptocurrency transactions in the votes instead of the blocks~\cite{KellerGBG23}.
This increases transaction throughput and enables faster confirmations.
However, there are now two types of confirmations:
\emph{block confirmations} span across epochs and retain the consistency guarantees of parallel proof-of-work, while the faster \emph{vote confirmations} within an individual epoch do not possess these guarantees.
The focus of this paper, however, is not on consistency and throughput, but on incentives.
In this regard, Keller et al.~\cite{KellerGBG23} introduce a promising reward scheme which actively discourages non-linearities in the blockchain.

Recall that sequential proof-of-work assigns a constant amount of reward to each block \emph{in the chain}.
Orphans do not receive a reward.
In that sense, sequential proof-of-work punishes non-linear chains by handing out fewer rewards.
This is natural, as linearizing the blocks and the transactions contained therein is the main objective of the system.
Unfortunately, this punishment is also inherently unfair.

Consider a situation where two miners, one weak with 2\,\% of the hashrate and one strong with 20\,\% of the hashrate, mine blocks around the same time, unintentionally creating a fork.
At most one of the blocks fits into the chain, so one miner will come away empty-handed.
Naturally, both miners will try to confirm their own block.
Assuming the other miners %
are unbiased, 38\,\% of the hashrate supports the weak miner's block and 62\,\% of the hashrate supports the strong miner's block.
Sequential proof-of-work punishes non-linearity, but does so in favor of the stronger miners.

\begin{figure}
  \centering
  \input{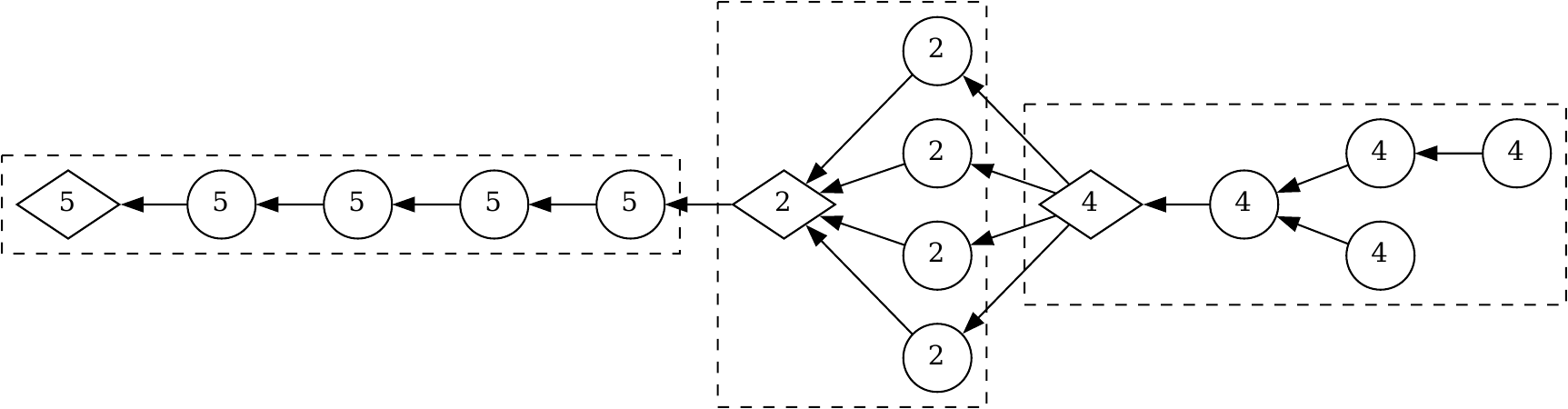}
  \caption{
    Example chain for parallel proof-of-work with tree-style voting and $k = 5$ proofs-of-work per block.
    Diamonds represent blocks and circles represent votes.
    Arrows indicate hash-references.
    Dashed boxes mark epochs.
    The labels show the discount incentive scheme where rewards are proportional to the depth of the vote tree.
  }\label{fig:tree}
\end{figure}

Tree-style voting addresses this issue by ensuring all involved miners are punished equally.
Within an epoch, every proof-of-work receives the same amount of reward.
The size of the reward, however, is scaled proportionally to the depth of the epoch's vote tree.
Figure~\ref{fig:tree} shows an example blockchain for tree-style voting with the discounted rewards on the labels.
This approach largely eliminates the bias against weak miners in benign networks~\cite{KellerGBG23}.
At the same time, the overall punishment---minus one unit of reward per block off the longest chain---is the same as in sequential proof-of-work.

But it is also important to consider malicious behaviour.
Recall that sequential proof-of-work suffers from an incentive attack, known as Selfish Mining~\cite{EyalS14, SapirshteinSZ16}.
The overall idea is as follows:
the attacker withholds his own blocks, deliberately causing and growing a fork.
Now, whenever the defenders mine a block, the attacker can orphan this block by releasing a longer (or matching) chain.
In the long run, after the difficulty is adjusted in response to all the orphans caused, the attacker reaps more rewards.
The attack proves to be effective for attackers who control more than a third of the hashrate, but depending on network conditions, the threshold is even lower~\cite{EyalS14}.
Tree-style voting mitigates this kind of attack by punishing information withholding.
Withholding votes implies forking the tree and consequently results in lower rewards.
This defense is effective, both in comparison to sequential proof-of-work and plain parallel proof-of-work~\cite{KellerGBG23}.

We summarize that tree-style voting, compared to plain parallel proof-of-work, provides faster first confirmations, higher throughput, and is less prone to incentive attacks.
It punishes non-linear chains like sequential proof-of-work, but does so in a fair manner.
These achievements are significant; still, we must ask the question, \enquote{Why limit it to a tree?}

\subsection{Parallel Proof-of-Work with DAG-Style Voting}\label{sec:dag}

\begin{figure*}
  \centering
  \input{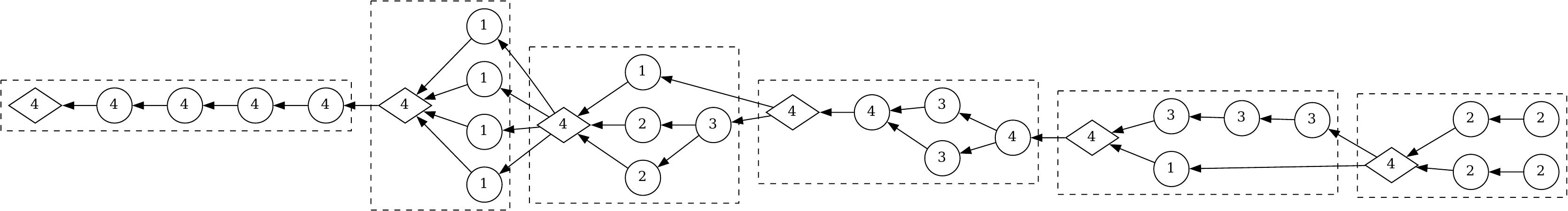}
  \caption{
    Example chain for parallel proof-of-work with DAG-style voting and $k = 5$ proofs-of-work per block.
    Arrows indicate hash-references.
    Dashed boxes mark epochs.
    The labels show the targeted discount incentive scheme where rewards are proportional to the number of ancestors and descendants within the same epoch.
  }\label{fig:dag}
\end{figure*}

We propose a new variant of parallel proof-of-work that structures the epochs' votes as directed acyclic graph (DAG).
In our protocol, votes can have \emph{multiple} parents in the same epoch.
When appending a new vote, miners refer to all leaves they know.
The other consensus rules remain unchanged.

Tree-style voting instructs the miners to build a linear chain when possible and allows them to branch out when needed.
Unfortunately, once a fork is created, it persists until the end of the epoch.
In contrast, with DAG-style voting miners join all existing forks back into a linear chain as soon as possible.
The protocol records the votes' temporal relationships more precisely on chain.
In extension, DAG-style voting can better order the transactions contained within the votes.

In principle, the tree-style discount scheme presented above could be applied to DAG-style voting without changes.
However, we question a core aspect of its design: the idea that everyone involved in a non-linear epoch is punished equally.
Consider a situation where all but the last two miners of an epoch coordinate perfectly and build a single linear branch.
Then, at the end of the epoch, two badly connected miners accidentally produce two votes forking the chain, also triggering the transition to the next epoch.
In such a scenario, the reward is reduced for all miners, despite the fact that the two responsible miners can be clearly identified, considering only the information recorded on chain.

Our DAG-style discount scheme mitigates this problem.
The core principle is not to reward the proof-of-work itself, but rather the amount of linearity it contributes to the chain.
At the end of each epoch, for each vote, we count the vote's ancestors and descendants within the epoch, setting the reward proportionally.
More siblings imply less direct lineage, which in turn implies a lower reward.
Figure~\ref{fig:dag} shows an example chain for parallel proof-of-work with DAG-style voting.
The rewards of the proposed discount scheme are depicted on the labels.
Note that, unlike before, the punishment for non-linearities is now targeted at the offending votes.

\section{Evaluation}\label{sec:eval}

We claim that DAG-style voting improves on tree-style voting in terms of resilience against incentive attacks, and does so without sacrificing the other desired properties.
We will support our claim with an elaborate analysis, after briefly covering the preservation of the other properties.

\paragraph{Consistency}
Keller and Böhme~\cite{KellerB22} argue that parallel proof-of-work yields higher consistency guarantees than sequential proof-of-work.
The protocol dictates that miners extend the longest chain of blocks and, in cases where there are two competing blockchains of equal length, vote for the one which has received more votes already.

DAG-style voting applies the same rules and hence inherits the consistency guarantees of parallel proof-of-work.
The same argument has been made for tree-style voting~\cite{KellerGBG23}.

\paragraph{Fast First Confirmations and Throughput}
The original version of parallel proof-of-work records cryptocurrency transactions in the block~\cite{KellerB22}.
Keller et al. propose with Tailstorm to record the cryptocurrency transactions in the votes instead~\cite{KellerGBG23}.
Leveraging the idea that more frequent small blocks cause smaller delays than less frequent large blocks~\cite{Rizun16}, they argue that overall throughput can be increased.
Additionally, Tailstorm employs tree-style voting, which records some temporal relationships between votes on the chain.
This yields faster first confirmations, although we want to highlight that the vote confirmations do not provide the same consistency guarantees as full block confirmations.

DAG-style voting increases the number of references between votes, thereby potentially providing more vote confirmations, never less.
The throughput argument, on the other hand, applies equally to all variants of parallel proof-of-work.

\paragraph{Fairness}
Sequential proof-of-work exhibits a bias against weak miners, as their blocks are more regularly orphaned~\cite{AlzayatMCGL21, KellerGBG23}.
This can be mitigated by avoiding orphans and ensuring that all blocks receive a fair reward~\cite{KellerGBG23}, which is characteristic for all variants of parallel proof-of-work.

\paragraph{Incentive Attacks}

Tailstorm leverages tree-style reward discounting to punish block withholding and thereby mitigate incentive attacks~\cite{KellerGBG23}.
The authors provide evidence by searching for effective attack strategies using reinforcement learning.

In the following, we recapitulate their analytic framework and apply it to all four protocols described in Section~\ref{sec:proto}.
Our results confirm the intuition provided above.
We will find out that DAG-style voting is the most resilient to incentive attacks under all considered network conditions.

\subsection{Attack Model}\label{sec:attack}

Recall our system model from Section~\ref{sec:model}.
Briefly, the system consists of multiple nodes that communicate over a peer-to-peer network.
Defenders instruct their nodes to adhere to the protocol, while malicious nodes may not.

We focus on worst-case scenarios, thus assuming all attackers collaborate.
Hereafter, we refer to the group of attackers as a single entity, the attacker.
Following the same reasoning, we consider only one malicious node.

\paragraph{Action Space}\label{par:actions}
Closely following the literature on Selfish Mining~\cite{EyalS14,SapirshteinSZ16,ZurET20}, we limit the attacker's capabilities.
We focus on attacks where the attacker forks the chain and withholds their blocks and votes.
Throughout our analysis, we assume the attacker maintains at most one private fork.
Upon learning about a new block or vote, whether mined locally or received from the network, the attacker can choose from a set of predefined actions.
There are two types of actions that enable the attacker to separately control \emph{withholding and forking} and \emph{inclusion of votes}.
The actions for \emph{withholding and forking} mirror those defined by Sapirshtein et al.~\cite{SapirshteinSZ16} to find optimal Selfish Mining strategies in Bitcoin:

\begin{enumerate}
  \item \emph{Wait}.\label{act:wait} Continue to mine on the private chain, withholding new blocks and votes.
  \item \emph{Match}.\label{act:match} Release just enough blocks and votes in topological order, to make the defenders indifferent between their chain and the attacker's chain.
  \item \emph{Override}.\label{act:override} Release just enough blocks and votes in topological order, to make the defenders prefer the attacker's chain over their own.
  \item \emph{Adopt}.\label{act:adopt} Discard the private chain and begin a new fork after the latest defender block.
\end{enumerate}

The above actions are tailored for sequential proof-of-work.
But parallel proof-of-work and its variants, have one additional degree of freedom:
the attacker can choose which votes to include when assembling a new block or, in DAG-style voting, when appending a new vote.
The actions for the \emph{inclusion of votes} have been proposed for the analysis of Tailstorm~\cite{KellerGBG23}:

\begin{enumerate}[i)]
  \item \emph{Inclusive}.\label{act:incl} Consider all available votes.
  \item \emph{Exclusive}.\label{act:excl} Consider only the attacker's votes.
\end{enumerate}

For all four protocols, honest behavior can be replicated by always considering all votes (\emph{Inclusive} action), releasing one's own blocks when they are fresh (\emph{Override} action), and following the defenders' chain when they mine a new block (\emph{Adopt} action).

\paragraph{Parameters}
The efficacy of any attack using the specified actions relies on two parameters.
The first is the attacker's relative hashrate, denoted as $\alpha$.
The second parameter reflects the attacker's network capabilities.
Notably, in \emph{sequential} proof-of-work, a node encountering two competing chains of the same length continues mining to confirm the first received block.
Similar situations occur in \emph{parallel} proof-of-work protocols, for example, when two chains have identical lengths and the same number of votes.
Attacker's can exploit this in practice by delaying and reordering messages.
In our model, attackers deliberately create such scenarios using the \emph{Match} action.
The $\gamma$ parameter defines the fraction of defenders deceived by this tactic, specifically those who continue mining on the attacker's block instead of the defenders' chain.
Our definitions of $\alpha$ and $\gamma$ closely align with prior work on Selfish Mining~\cite{EyalS14,SapirshteinSZ16, ZurET20, KellerGBG23}. %

\paragraph{Revenue}
We assume the attacker evaluates her utility in terms of the cryptocurrency itself, or equivalently, under a fixed exchange rate.
We also assume that the attacker incurs a constant fixed cost to maintain the mining operation over time.
The attacker's goal is to maximize the reward received per unit of time.
When doing so, it is essential to consider the time horizon of the attack.

In the short term, the attacker can maximize her revenue by adhering to the protocol.
This is because honest behavior results in a linear chain where all mined blocks receive the maximum reward.
Securing full reward for all blocks mined is optimal, provided that the attacker's mining rate remains unchanged, which is true for the short term.

In the long term, however, the mining rate can change due to the difficulty adjustment algorithm's inability to account for orphaned blocks or votes.
Creating orphans initially slows down the chain but, as soon as the difficulty adjustment takes effect, the target rate is restored.
An attacker who causes orphans among the defenders' blocks and votes can ultimately include more of her own blocks per time.

To account for the effects of the difficulty adjustment algorithm, we assume that the difficulty has already adapted to the attack.
In other words, we assume the longest chain grows at a constant rate.
In our analysis, we optimize for \emph{normalized revenue}, defined as the amount of reward received, relative to the hypothetical reward that would be minted in a strictly linear chain.
It is important to note that the simpler \emph{relative reward} metric, often used for Selfish Mining against Bitcoin, does not adequately capture the impact of reward discounting in tree- and DAG-style voting.

\subsection{Reinforcement Learning}\label{sec:rl}

We employ reinforcement learning to identify effective attack strategies against the four protocols outlined in Section~\ref{sec:proto}.
This approach was originally proposed by Hou et al.~\cite{squirrl}.
We reuse our learning pipeline, as presented earlier to analyze Tailstorm~\cite{KellerGBG23}.
We provide a summary here and make the complete implementation available online~\cite{trainingrepo}.
For this paper, we added specifications for the three parallel proof-of-work protocols described in Sections~\ref{sec:par} to~\ref{sec:dag}.

\paragraph{Terminology}
The \emph{agent} makes decisions and learns through its interactions with an environment.
The \emph{environment} is the domain or context the agent operates in, providing the conditions and rules that defines the optimization problem.
A \emph{step} represents a single instance of interaction between the agent and the environment, during which the agent makes an \emph{observation} of the environment, performs an \emph{action}, and receives a \emph{reward} as feedback.
An \emph{episode} represents a continuous flow of interaction between the agent and the environment.
The entire learning process, consisting of multiple episodes, is referred to as a \emph{training run}.
The algorithmic representation of the agent has adjustable parameters that control its behaviour and which we call \emph{weights}.
During a training run, the weights are continuously updated to maximize the reward received per episode.
A \emph{policy} is one set of weights or, equivalently, one specific mapping from observations to actions.

\paragraph{Environment}

We deploy 20 defender nodes and one malicious node in a simulated network environment.
The agent controls the malicious node.
Whenever the agent learns about a new block, she chooses one of the actions defined in Paragraph~\ref{par:actions}.
The decision is based on the following observed variables:
\begin{enumerate}
  \item the number of blocks in the defenders' public branch,
  \item the number of blocks in the attacker's private branch,
  \item the number of votes confirming the last block in the defenders' public branch,
  \item the total number of votes confirming the last block in the attacker's private branch,
  \item the number of attacker votes confirming the last block in the attacker's private branch, and
  \item whether the new block was mined locally or received from the network.
\end{enumerate}

We set the malicious node's relative hashrate $\alpha$ and allocate the rest evenly among the defending nodes.
During the simulation, we choose individual network delays to reproduce the communication advantage $\gamma$.

We evaluate all four protocols defined in Section~\ref{sec:proto}:
sequential and plain parallel proof-of-work with constant reward scheme and parallel tree-style and DAG-style voting with their respective discount reward schemes.
The parallel protocols all use $k = 8$ proofs-of-work per block to enable comparison with the existing results for Tailstorm~\cite{KellerGBG23}.
For each protocol, we consider 15 different network scenarios:
all combinations of $\gamma \in \{0.05, 0.5, 0.95\}$ and $\alpha \in \{0.25, 0.3, 0.35, 0.4, 0.45\}$.
Overall, this makes 60 different training tasks.

\begin{figure*}
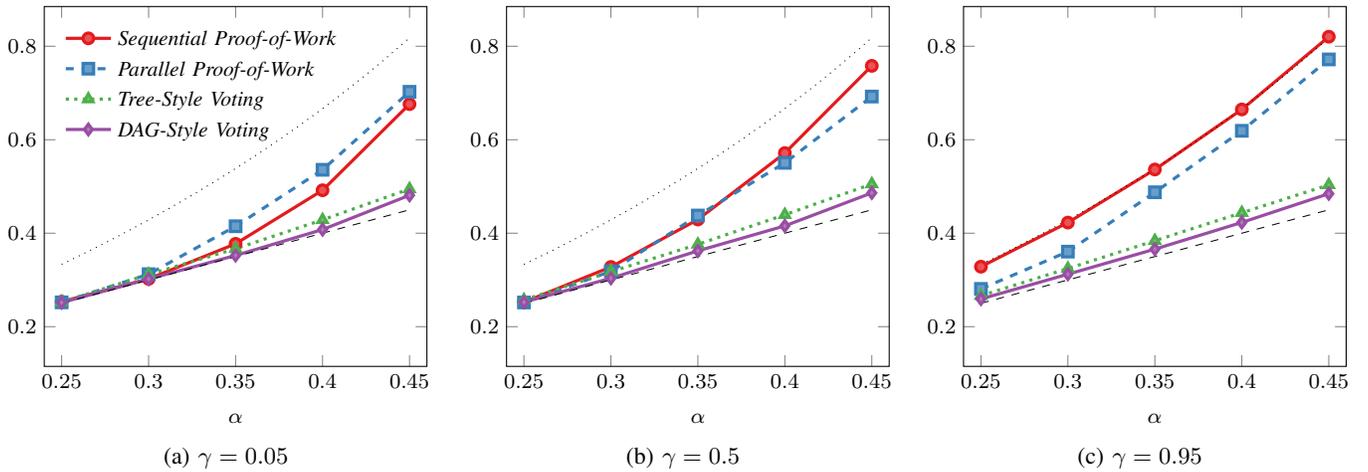

  \begin{subfigure}{.33\linewidth}
    \resultaxis{
      \addplot+ table [x=alpha, y=gamma05_seq_rl, col sep=comma] {data/rpp.csv};
      \addlegendentry {\emph{Sequential Proof-of-Work}}
      \addplot+ table [x=alpha, y=gamma05_par8c_rl, col sep=comma] {data/rpp.csv};
      \addlegendentry {\emph{Parallel Proof-of-Work}}
      \addplot+ table [x=alpha, y=gamma05_tree8d_rl, col sep=comma] {data/rpp.csv};
      \addlegendentry {\emph{Tree-Style Voting}}
      \addplot+ table [x=alpha, y=gamma05_dag8d_rl, col sep=comma] {data/rpp.csv};
      \addlegendentry {\emph{DAG-Style Voting}}
    }
    \subcaption{$\gamma=0.05$}\label{fig:rlres:g05}
  \end{subfigure}\hfill
  \begin{subfigure}{.33\linewidth}
    \hfill
    \resultaxis{
      \addplot+ table [x=alpha, y=gamma50_seq_rl, col sep=comma] {data/rpp.csv};
      \addplot+ table [x=alpha, y=gamma50_par8c_rl, col sep=comma] {data/rpp.csv};
      \addplot+ table [x=alpha, y=gamma50_tree8d_rl, col sep=comma] {data/rpp.csv};
      \addplot+ table [x=alpha, y=gamma50_dag8d_rl, col sep=comma] {data/rpp.csv};
    }
    \subcaption{$\gamma=0.5$}\label{fig:rlres:g50}
  \end{subfigure}\hfill
  \begin{subfigure}{.33\linewidth}
    \hfill
    \resultaxis{
      \addplot+ table [x=alpha, y=gamma95_seq_rl, col sep=comma] {data/rpp.csv};
      \addplot+ table [x=alpha, y=gamma95_par8c_rl, col sep=comma] {data/rpp.csv};
      \addplot+ table [x=alpha, y=gamma95_tree8d_rl, col sep=comma] {data/rpp.csv};
      \addplot+ table [x=alpha, y=gamma95_dag8d_rl, col sep=comma] {data/rpp.csv};
    }
    \subcaption{$\gamma=0.95$}\label{fig:rlres:g95}
  \end{subfigure}%
  \caption{%
    Revenue obtained from reinforcement learning based attacks.
    The block race advantage $\gamma$ is represented on the facet.
    The $x$-axis shows the attacker's relative hashrate $\alpha$.
    The $y$-axis shows the normalized long-term revenue obtained by applying the best attack found.
    The different protocols and reward schemes are represented by the curves' style and color.
    The black dashed line shows the expected revenue for honest behavior, $y = \alpha$.
    The black dotted line shows $y = \alpha / (1-\alpha)$ which represents an upper bound in sequential proof-of-work~\cite{SapirshteinSZ16}.
  }\label{fig:rlres}
\end{figure*}

\paragraph{Training}

We solve each training task individually, using an established reinforcement learning algorithm, proximal policy optimization (PPO)~\cite{ppopaper,sb3}.
For each task, we setup six training runs using different hyperparameters:
all combinations of three entropy coefficients (0.01, 0.005, 0.001) and two learning rates (0.0003 and 0.001).
The other hyperparameters are shared across all runs:
we use neural networks with three dense layers of $2^8$ neurons each, discount factor~1, minibatch size $2^{10}$, batch size $2^{17}$, and stop training after 30 million steps.
We stop individual episodes as soon as the longest chain contains $2^7$ proofs-of-work.

Overall, we conduct 360 training runs.
Each training run takes about 9.5 hours on a single core of a 2022 AMD processor (Zen 3, Ryzen 5 5600).
This amounts to 142.5 days of training or, parallelized across all 6 cores, about 24 days.

During each run, we select and store the last weights and the ones yielding the best results on intermediate evaluations.
This results in 720 policies, twelve for each training task.

\paragraph{Best Available Policies}

We conduct 100 independent simulations for each policy, stopping each simulation after $2^{11}$ proof-of-work puzzles are solved.
For each training task, we select the policy yielding the highest mean revenue as the best available policy.
This result in 60 selected policies, one per combination of network scenario and protocol.

\subsection{Results}\label{sec:results}

Figure~\ref{fig:rlres} shows our results.
The network parameter~$\gamma$ is represented on the facet, the relative hashrate $\alpha$ on the $x$-axis.
The $y$-axis shows the mean normalized revenue obtained from following the best available policy for the scenario at hand.
We draw four curves with different colors and styles, one for each protocol.
As reference, we include a line for honest behaviour, $y = \alpha$, and the curve $y = \alpha / (1 - \alpha)$ which represents an upper bound in sequential proof-of-work~\cite{SapirshteinSZ16}.

Our results support three main conclusions.
First, plain parallel proof-of-work with constant rewards (red curve, Sec.\,\ref{sec:par}), for low $\gamma$, is less resilient to incentive attacks than sequential proof-of-work (blue curve, Sec.\,\ref{sec:seq}).
This confirms the findings of Keller et al.~\cite{KellerGBG23}, who assessed the AFT\,'22 version of parallel proof-of-work~\cite{KellerB22}.
However, it challenges their assertion that eliminating the leader election mechanism mitigates this issue.
We remove the leader election mechanism in our version of parallel proof-of-work (see Sec.\,\ref{sec:par}), but we observe that this alteration does not resolve the problem.
Since Keller et al.\ did not directly evaluate parallel proof-of-work without leader election, but rather a version of tree-style voting with constant reward, we place greater confidence in our findings.

Second, we confirm that reward discounting mitigates the problem.
Both tree-style (green curve, Sec.\,\ref{sec:tree}) and DAG-style reward discounting (purple curve, Sec.\,\ref{sec:dag}) exhibit greater resilience to incentive attacks than sequential proof-of-work under all analyzed network conditions.

Third, our proposed DAG-style reward discounting scheme shows more resilience to incentive attacks than the tree-style reward discounting originally proposed in Tailstorm~\cite{KellerGBG23}, consistently across all analyzed network conditions.

\section{Protocol Configuration}\label{sec:config}

The original work on parallel proof-of-work~\cite{KellerB22} offers detailed guidance for configuring the protocol.
Here, we present a single configuration optimized for a particular worst-case scenario:
we assume the attacker controls at most one quarter of the mining power, and that messages propagate within 2 seconds.
Aiming for a block interval of ten minutes, akin to Bitcoin, Keller and Böhme recommend using $k = 51$ proofs-of-work per block to maximize consistency.
Under these assumptions, this configuration ensures consistency with a failure probability of only 0.02\,\% after one block confirmation, or approximately 10 minutes~\cite{KellerB22}.
For context, sequential proof-of-work's consistency failure probability is at most 1\,\% over the same duration~\cite{Guo022}---a factor 50 difference.
Since tree-style and DAG-style voting do not alter the consensus rules of parallel proof-of-work, we recommend the same settings for these protocols.
This configuration results in an expected vote interval of approximately 12 seconds, yielding a transaction throughput comparable to Ethereum proof-of-work.

\section{Discussion}\label{sec:discussion}

We use reinforcement learning to search reward optimizing attacks.
Observing the effectiveness of the uncovered attacks, we conclude that some protocols are more resilient to incentive attacks than others.
We want to emphasize that this approach depends on the effectiveness of the attack search itself.
In principle, the reinforcement learning algorithm might find effective attacks against one protocol but not the other, leading us to incorrect conclusions.
We inherit this limitation from the analysis of Tailstorm~\cite{KellerGBG23}.
Keller et al.~verify that the attacks found meet established optimality results for Bitcoin.
Additionally, the attacks obtained from reinforcement learning outperform their handcrafted attacks for all protocols.
This supports their conclusion that reinforcement learning is indeed effective.
As we closely follow their approach, using the same assumptions, attack space, and reinforcement learning algorithm, our results are as reliable as theirs.

Nevertheless, we acknowledge that finding attacks with Markov Decision Processes (MDPs) and exact solving techniques~\cite{SapirshteinSZ16, GervaisKWGRC16, ZhangP19, ZurET20} would yield more dependable results.
Unfortunately, complex blockchain protocols, like tree-style and DAG-style voting, lead to a state space explosion which, so far, has rendered these techniques impractical.

On a separate note, we want to highlight that we here have incrementally transformed parallel proof-of-work into a DAG-protocol.
The gap to other DAG-protocols like PHANTOM~\cite{SompolinskyWZ21} or DAGKNIGHT~\cite{SompolinskyS22} appears relatively small, particularly when DAG-style voting is configured with many proofs-of-work per block.
For future work, we plan to evaluate whether our attacks against parallel proof-of-work without discounting apply to these other protocols, and whether our DAG-style reward discounting provides a suitable mitigation.

Lastly, we want to spotlight a promising avenue for future research: both tree-style and our DAG-style reward discounting have been designed in an ad-hoc manner.
Researchers in mechanism design might be able to find an \emph{optimal} reward scheme that is \emph{most resilient} to incentive attacks.

\section{Conclusion}\label{sec:conclusion}

This work builds upon several recent advancements in proof-of-work consensus protocol design.
We proposed a new protocol, parallel proof-of-work with DAG-style voting, which integrates the consistency of parallel proof-of-work as well as the high transaction throughput and low confirmation latency of Tailstorm with a new, more targeted reward discounting scheme.
Along the way, we systematized the key design decisions of parallel proof-of-work and tree-style voting.
We uncovered new incentive attacks against parallel proof-of-work without reward discounting.
Targeted discounting mitigates this issue more effectively than earlier proposals.
Given the resemblance of our protocol to other DAG-protocols, some deployed in practice, our findings warrant future research into the applicability of the found attacks and proposed mitigation.

\bibliographystyle{IEEEtran}
\bibliography{dblp,manual}

\end{document}